\documentclass[a4paper]{article}

\usepackage{INTERSPEECH2022}

\title{Unsupervised Word Segmentation using K Nearest Neighbors}
\name{Tzeviya Sylvia Fuchs$^1$, Yedid Hoshen$^2$ and Joseph Keshet$^1$}
\address{
  $^1$ Bar-Ilan University, Ramat-Gan, Israel \\ 
  $^2$ The Hebrew University of Jerusalem}
\email{fuchstz@cs.biu.ac.il, yedid.hoshen@mail.huji.ac.il, jkeshet@cs.biu.ac.il}

\usepackage{xcolor}

 \newfont{\msym}{msbm10}

\usepackage{algorithmic}

\newsavebox{\ieeealgbox}

\begin{document}

\maketitle
\begin{abstract}
  In this paper, we propose an unsupervised kNN-based approach for word segmentation in speech utterances. Our method relies on self-supervised pre-trained speech representations, and compares each audio segment of a given utterance to its K nearest neighbors within the training set. Our main assumption is that a segment containing more than one word would occur less often than a segment containing a single word. Our method does not require phoneme discovery and is able to operate directly on pre-trained audio representations. This is in contrast to current methods that use a two-stage approach; first detecting the phonemes in the utterance and then detecting word-boundaries according to statistics calculated on phoneme patterns. Experiments on two datasets demonstrate improved results over previous single-stage methods and competitive results on state-of-the-art two-stage methods.

\end{abstract}

\section{Introduction}

Models for unsupervised word segmentation on unlabelled speech data can enable language acquisition in low-resource languages \cite{jansen2013summary}. They can also shed new light on human cognition \cite{rasanen2012computational, dupoux2018cognitive}. The task of word segmentation is to predict all word boundaries within an utterance \cite{kamper2017embedded, kamper2017segmental, shain2020acquiring, rasanen2020unsupervised, kamper2020towards, bhati2021segmental}. Word segmentation is often the first stage, followed by a subsequent clustering stage for unsupervised word-like unit discovery. Although word segmentation is usually a trivial task in written text, it is very challenging for speech as there is typically no spacing between uttered words. Additionally, no two utterances of the same word sound the same. This task has been attempted for decades but results are still far from satisfactory. 

Previous works can be divided into two main approaches: single-stage and two-stage methods. Two-stage word-segmentation methods first attempt to perform unsupervised phoneme segmentation and classification. Such methods then utilize the discrete phonetic representation of the utterance to perform word segmentation. Many current approaches use a ``language model''-like method for the phonemes, which attempts to estimate the likelihood of the subsequence of phonemes. This can be performed by classical natural language processing approaches such as Adapter Grammar (AG) \cite{johnson2009improving, kamper2020towards}. Other methods attempt to learn deep neural prediction models \cite{bhati2021segmental, cuervo2021contrastive}, also inspired by more recent natural language processing literature. Two-stage methods currently lead unsupervised word segmentation benchmarks - most notably, the Buckeye dataset, which is commonly used. While at present, two-stage approaches enjoy strong performance, they have significant limitations. The most notable ones are: i) the requirement for two sets of models - using both frame and phoneme level inputs. ii) Using an intermediate phoneme discovery task, which may result in a compounding of errors.

Single-stage word segmentation methods differ from two-stage methods by not requiring a phoneme segmentation stage. Instead, they directly attempt to perform word segmentation using the frame-level features. This approach is attractive, as it does not suffer from the two limitations of two-stage methods mentioned above. Despite these properties, single-stage techniques are not typically used by recent methods. A notable previous single-stage method is Embedded Segmental K-means (ES-KMeans) \cite{kamper2017embedded}. Although an inspiring approach, ES-KMeans underperforms recent two-stage approaches.

Here, we first propose a modernized version of ES-KMeans that uses strong pertrained features. We show that replacing the classical features by recent, deep features can significantly boost unsupervised clustering performance. Nonetheless, this approach does not scale to large-scale datasets such as Buckeye. The reason is that the very large vocabulary of these datasets does not match clustering assumptions well enough to inform the segmentation. Instead, we propose an alternative approach that we name Deep Segmental-kNN. Our approach hypothesizes that segments of the utterance at the boundary of two words, experience higher variability (and therefore lower likelihood) than intra-word segments. Although these ideas are also used by many two-stage approaches - our approach differs by not requiring phoneme information to calculate the likelihood scores. We utilize recent results from the anomaly detection literature \cite{reiss2021panda} that showed that simple kNN anomaly detectors coupled with strong pretrained deep features can achieve outstanding performance. Specifically, we extract deep features from short segments of the utterance and compare them to the features of their nearest neighbors in the training set. Segments with large anomaly scores suggest a word boundary. Using a peak-finding stage on the anomaly score of the test utterance, the word boundary segmentation is recovered . 

We perform experiments on two datasets: YOHO \cite{campbell1995testing} and Buckeye \cite{pitt2005buckeye} demonstrating the effectiveness of our approach. We compare our method to previous works on the Buckeye dataset. Our method, which is far simpler than previous methods, achieves performance comparable to the state-of-the-art.

\section{Method}

The objective of our method is to take a speech signal as input and return a set of segments, each containing a single word. We denote the speech signal as $\bar{\mathbf{x}} = [x_1,x_2, \ldots, x_T]$, where $x_t$ denotes the sample at time $t$. The input signal $\bar{\mathbf{x}}$ could be represented using some feature extractor $\phi$. The signal $\bar{\mathbf{x}}$ is encoded into a feature vector sequence $\bar{\mathbf{f}} = [\mathbf{f}_1,\mathbf{f}_2 \ldots, \mathbf{f}_N]$ such that $\bar{\mathbf{f}} = \phi(\bar{\mathbf{x}})$. The segments are described by a sequence of boundaries $\bar{\mathbf{b}} = [1,b_1, \ldots, b_M,T]$, where each boundary $b_m$ denotes a particular time. Ideally, $\hat{\mathbf{x}}^m = [x_{b_{m-1}},x_{b_{m-1}+1}, \ldots,x_{b_{m}}]$ should correspond to the segment of the utterance that corresponds to word number $m$. $\hat{\mathbf{f}}^m$ would be its feature representation. Word segmentation is both useful on its own, or is used as a preliminary step for subsequent word clustering.

\subsection{Preliminary Approach: Deep Features for ES-KMeans}

Embedded Segmental KMeans \cite{kamper2017embedded} (ES-KMeans) is a popular single-stage word segmentation approach. To infer the word boundaries $\bar{\mathbf{b}}$ from signal $\bar{\mathbf{x}}$, ES-KMeans relies on two key assumptions: i) the features of a single word are more similar to each other than the features of other words. ii) precisely segmented words can be more compactly clustered into a small number of means than a poorly segmented utterance consisting of combinations of words. Unfortunately, the first assumption is not always satisfied when using classical features such as MFCCs utilized by ES-KMeans. Although ES-KMeans is not competitive with the current state-of-the-art, it may be speculated that this is due to its use of classical MFCC features. A reasonable hypothesis is that strong pretrained features will satisfy this assumption with higher fidelity as they are more invariant to nuisance factors. 

We perform experiments to investigate two questions: i) are deep features superior to classical features for unsupervised word clustering? ii) would this higher clustering ability improve the word segmentation quality of ES-KMeans?. To answer the first question, we segmented the signal of each word in the YOHO and Buckeye-dev datasets using the \textit{ground truth} labels. We then extracted both MFCC and deep features using the Wav2Vec 2.0 encoder pretrained on the LibriSpeech dataset in a fully self-supervised manner (i.e. without using any labels). Clustering is performed by running KMeans on the extracted features using the ground truth number of words. The results are evaluated using the commonly reported accuracy (ACC) metric which assigns each cluster to a unique word using the Hungarian bipartite matching algorithm \cite{kuhn1955hungarian, munkres1957algorithms}. We indeed find that deep features are significantly better at clustering than classical features. The ACC metric for MFCC is $26.0\%$ (YOHO) and $18.5\%$ (Buckeye) while for deep features the ACC is $76.5\%$ (YOHO) and $28.9\%$ (Buckeye). However, despite the strong clustering performance, we did not find that deep features improved the results of ES-KMeans on the large vocabulary Buckeye dataset (F1-score:$19.1$) although they did help achieve acceptable performance on the YOHO dataset (F1-score:$35.1$).

\subsection{Our Approach: Deep Segmental KNN}

We propose a new approach that utilizes anomaly detection for discovering word boundaries $\bar{\mathbf{b}}$ given signal $\bar{\mathbf{x}}$. The main principle behind our approach is that intra-word segments have higher likelihood (i.e. a particular segment appears with higher frequency in the dataset) than segments that include multiple words. This is sensible as single words occur more frequently than word pairs. The main technical challenge is the probability estimation of the signal segment. Formally, given a feature representation $\hat{\mathbf{f}}^m$ (of segment $m$ in the speech signal), we wish to learn a probability density function $p$ that takes in the signal segment and returns its probability density $p(\hat{\mathbf{f}}^m)$.

This challenge has been tackled by two-stage methods, which first attempted to discover phonemes in the signal. By representing the signal as a sequence of phonemes, they then estimate $p(\hat{\mathbf{f}}^m)$ using different approaches. One approach is using N-Grams as a language model, but since the discovered phonemes are noisy and often inconsistent, such models frequently fail. Recent approaches train models to predict the next phoneme segments given the previous phonemes. This however still requires phoneme discovery.

Here, we propose a new approach that does not require phoneme discovery. Our approach, Deep Segmental KNN (DSegKNN), proposes to estimate $p(\hat{\mathbf{f}}^m)$ directly using non-parametric estimation. We utilize recent results from the anomaly detection literature, showing that the kNN estimators can be highly effective for discovering anomalous (low-likelihood) patterns. Reiss et al. \cite{reiss2021panda} discovered that such methods achieve the state-of-the-art anomaly detection tasks when coupled with strong pretrained representations. Strong representations are required as kNN is highly sensitive to the similarity function used. It is therefore necessary to use a feature representation which is correlated to semantic similarity. 

\begin{figure}[t]
  \centering
  \includegraphics[width=\linewidth]{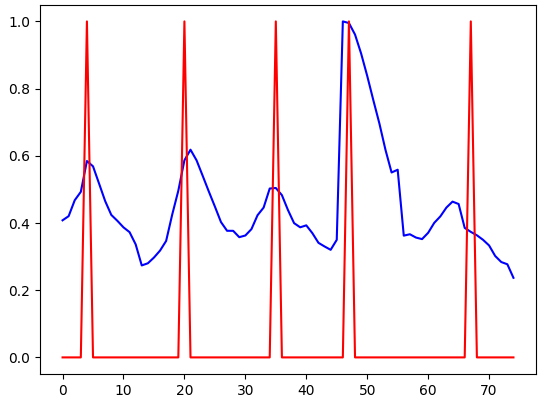}
  \caption{A test example from the YOHO dataset. The \textit{x} axis denotes time. The red bars are the ground truth boundaries and the blue graph is the anomaly score $\bar{\mathbf{s}}$. The peaks in $\bar{\mathbf{s}}$ roughly align with the ground truth boundaries.}
  \label{fig:knn_demonstration}
\end{figure}

DSegkNN first represents the input signal $\bar{\mathbf{x}}$ using a deep self-supervised trained feature extractor $\phi$. As explained above, the quality of the features is critical. We examined both Wav2Vec 2.0 and HuBERT encoders (feature extractors), which were trained in a fully self-supervised manner. Given the input signal's feature representation $\bar{\mathbf{f}}=\phi(\bar{\mathbf{x}})$, for each time point we extract a subsequence of features corresponding to a window of length $win$ centered at that time. We denote the feature subsequence at point $m$ as $\hat{\mathbf{f}}^m = [\mathbf{f}_{m - \frac{win}{2}}, \ldots, \mathbf{f}_m, \ldots, \mathbf{f}_{m + \frac{win}{2}}]$. The window-level feature subsequence is formed by concatenating all the feature representations computed across the sequence and flattening them, resulting in a vector of length $win * d_e$. Here $d_e$ denotes the dimension of each feature $\mathbf{f}_i$ extracted by encoder $\phi$ ($d_e$ is $768$ or $1024$ for Base or Large models in Wav2Vec 2.0). We similarly extract such window features for every time point in the training set. 

We define the anomaly criterion for each point $m$ by the $k$ nearest neighbor (kNN) distance between its feature representation $\hat{\mathbf{f}}^m$ and all the features in the training set. This can be formally written as:

\begin{equation}
    s_m = \sum_{\hat{\mathbf{f}} \in {\cal{F}}_{kNN}} \|\hat{\mathbf{f}} - \hat{\mathbf{f}}^m\|^2_2
\end{equation}

Where $s_m$ is the anomaly score at time $m$ and ${\cal{F}}_{kNN}$ denotes the set of $k$ feature vectors in the training set, such that their Euclidean distance to vector $\hat{\mathbf{f}}^m$ is minimal.

At the end of the anomaly detection stage, we obtain a temporal sequence of anomaly scores $\bar{\mathbf{s}} = \{s_1,s_2 \ldots s_N\}$. The task in the final stage is obtaining a set of word boundaries $\bar{\mathbf{b}}$ from the anomaly score sequence $\bar{\mathbf{s}}$. The main idea is that maximally anomalous times are most likely to correspond to word boundaries. We therefore utilize a peak detector to detect the most anomalous moments. One real example is presented in Fig.~\ref{fig:knn_demonstration}. The peak detector selects peaks that are no nearer than 100 ms apart (as few words are shorter than this duration). The output of our full method is a set of predicted boundaries $\bar{\mathbf{b}}$ for every input $\bar{\mathbf{x}}$.    

\section{Experimental Setup}

We evaluate our method on the Buckeye speech corpus \cite{pitt2005buckeye} and on the YOHO speaker verification dataset \cite{campbell1995testing}. 

\textbf{Buckeye \cite{pitt2005buckeye}.} The Buckeye corpus consists of conversational speech of $40$ speakers. We use the same experimental setup as in \cite{kreuk2020self}; we split the corpus at the speaker level using a ratio of $0.8$/$0.1$/$0.1$ for the train/val/test sets respectively. The long speech sequences are split into smaller segments by cutting during noise/silence instances, and each segment starts and ends with 20 ms of silence. In order to compare to previous works \cite{kamper2020towards, bhati2021segmental}, we evaluate on the development set, which consists of approximately 7 hours.

\textbf{YOHO \cite{campbell1995testing}.} The YOHO dataset contains 138 speakers, where each utterance contains a sequence of three two-digit numbers. The speakers were divided using a ratio of $0.8$/$0.1$/$0.1$ for the train/val/test sets respectively, and there are no overlapping speakers in the various data splits. Since the dataset is provided with no word alignments, the audio files were aligned to their given transcriptions using the Montreal Forced Aligner (MFA) \cite{mcauliffe2017montreal} for evaluation. Additionally, we applied a voice activity detector (VAD) on this dataset to detect silences in the beginnings and endings of each audio file, since the data consisted of long silences before and after the uttered phrase. Specifically, we used \texttt{py-webrtcvad}\footnote{\tiny{\texttt{https://github.com/wiseman/py-webrtcvad}}}. After removing opening and ending silences, the validation data consists of approximately 1 hour. The YOHO dataset represents the setting where datasets have a limited vocabulary.

\textbf{Metrics.} Similarly to \cite{kamper2020towards}, we report the precision, recall, F-score, over-segmentation (OS) and R-value. The OS metric evaluates how many more (or fewer) boundaries were predicted compared to the ground truth boundaries. The R-value \cite{rasanen2009improved} gives a perfect score ($1$) when a method has perfect recall ($1$) and perfect OS ($0$).

\subsection{Benchmark Evaluation}

\begin{table}[th]
  \caption{A comparison of word segmentation accuracy (\%) on the Buckeye validation set. Our method (DSegKNN) is compared to the state-of-the-art and older, established methods as reported in Bhati et al. \cite{bhati2021segmental}}
  \label{tab:buckeye_compare}
  \centering
  \resizebox{\columnwidth}{!}{
  \begin{tabular}{l c c c c c}
    \toprule
   \multicolumn{1}{c}{Model} & \multicolumn{1}{c}{\textbf{Prec.}} &  \multicolumn{1}{c}{\textbf{Recall}} &  \multicolumn{1}{c}{\textbf{F-score}} &  \multicolumn{1}{c}{\textbf{OS}}  &  \multicolumn{1}{c}{\textbf{R-val}}   \\
    \midrule
    ES-KMeans \cite{kamper2017embedded} &  30.7    &  18.0   & 22.7  & -41.2 & 39.7\\
    BES-GMM \cite{kamper2017segmental} &   31.7   &   13.8  &  19.2 & -56.6 & 37.9\\
    VQ-CPC DP \cite{kamper2020towards} &   15.5   &  \bf{81.0}   & 26.1 & 421.4  & -266.6\\
    VQ-VAE DP \cite{kamper2020towards} &     15.8  & 68.1 & 25.7  &  330.9  & -194.5 \\
    AG VQ-CPC DP \cite{kamper2020towards} &   18.2   &  54.1   &  27.3 & 196.4 & -86.5\\
    AG VQ-VAE DP \cite{kamper2020towards} &  16.4     &    56.8  & 25.5  &  245.2  & -126.5  \\
    Buckeye\_SCPC \cite{bhati2021segmental} &   \bf{35.0}   &  29.6  &  \bf{32.1}  & -15.4  & \bf{44.5} \\
    DSegKNN  &   30.9   &  32.0   & \bf{31.5} & \bf{3.46} & 40.7\\
     \bottomrule
  \end{tabular}}
\end{table}

We compare our method, DSegKNN, to new  state-of-the-art and older established methods. The evaluation is performed on the Buckeye validation set and is presented in Table ~\ref{tab:buckeye_compare}. The baseline numbers are reported from Bhati et al. \cite{bhati2021segmental}. We can observe that classical single-stage methods, ES-KMeans and BES-GMM, achieved very low recall. On the other hand, the vector-quantized (VQ) based approaches achieved low precision. Our method DSegKNN achieves comparable results to the top two-stage method SCPC (and significantly outperformed the other one-stage methods). 

\textbf{Implementation details.} We used the following hyperparameters: The number of neighbors is $k=20$, the window size is $win=10$. We used only $200$ randomly selected training utterances. This significantly improved our method's speed while only slightly affecting the accuracy. For peak detection, we used the standard peak detection function by the \texttt{scipy} package. Smoothing the anomaly scores $\bar{\mathbf{s}}$ using a Gaussian 1D filter slightly improved results on YOHO but was not used on Buckeye. We used HuBERT-large for Buckeye and Wav2Vec2-Base for YOHO, and in both cases the architecture was pre-trained on unlabeled LibriSpeech.

\begin{table}[th]
  \caption{The effect of architecture and layer choice on DSegKNN. Results are presented in ($\%$) on the YOHO validation set.}
  \label{tab:segknn_results_arc}
  \centering
  \resizebox{\columnwidth}{!}{
  \begin{tabular}{l c c c c c }
    \toprule
   \multicolumn{1}{c}{} & \multicolumn{1}{c}{\textbf{Arc}} &  \multicolumn{1}{c}{\textbf{Layer}} &  \multicolumn{1}{c}{\textbf{Prec.}} & \multicolumn{1}{c}{\textbf{Recall}} & \multicolumn{1}{c}{\textbf{F-score}}  \\
    \midrule
      &  base  \cite{baevski2020wav2vec}  &  3  & 28.0  &  29.7  &   28.9 \\
      &  base   \cite{baevski2020wav2vec} &  6  & \bf{39.2}  &  \bf{43.9}  &  \bf{41.4}  \\
      &  base  \cite{baevski2020wav2vec}  &  11  &  26.9 &  33.4  &   29.8 \\
      &  hubert\_base  \cite{hsu2021hubert} &  3  &  27.5  & 30.2   & 28.8   \\
      &  hubert\_base  \cite{hsu2021hubert} &  6  & 37.6  &  39.8  &   38.7  \\
      &  hubert\_base  \cite{hsu2021hubert} &  11  &  36.1 &  38.9  & 37.4   \\
      \bottomrule
  \end{tabular}}
\end{table}

\subsection{Ablation Study}
\label{subsection:ablation}

We ablate the key hyperparameters of our method, to gain a deeper understand of its performance factors. The experiments are performed on the YOHO validation dataset.

\textbf{Model architecture:} Both the model architecture and the particular layer used affect the feature representation and consequently the performance of kNN. Results for two different architectures (Wav2Vec 2.0 and HuBERT) and different layers (3, 6, 11) are presented in Table~\ref{tab:segknn_results_arc}. We can observe that middle layers (here the sixth layer) achieve the best results. This is intuitive as choosing layers that are too shallow will not include enough global context while choosing overly deep layers can lose locality information. 

\textbf{Window length.} The window length is an important hyperparameter of our method as choosing too short a window might not observe enough information over multiple words, even at the boundary. Windows that are too long might always include multiple words and will not discover boundaries. We ablated the choice of window sizes in Table ~\ref{tab:segknn_results_params}. We find that windows of length $10$ frames - i.e. $200$ ms achieve the best results.    

\textbf{Number of nearest neighbors.} We investigated the effect of selecting different numbers of nearest neighbors ($[1,5,10,20,30]$). The results are provided in Table~\ref{tab:segknn_results_params}. It can be observed that beyond $5$ nearest neighbors the precise number has little effect, if any. We expect however that choosing extremely large values of $k$ can adversely affect the result. 

\textbf{Aggregation.} We compared feature aggregation by concatenation or averaging. Table~\ref{tab:segknn_results_params} shows that concatenation is far superior to averaging. This is expected as concatenation preserves the order of the features, which is lost by averaging. 

\begin{figure}[t]
  \centering
  \includegraphics[width=.95\linewidth]{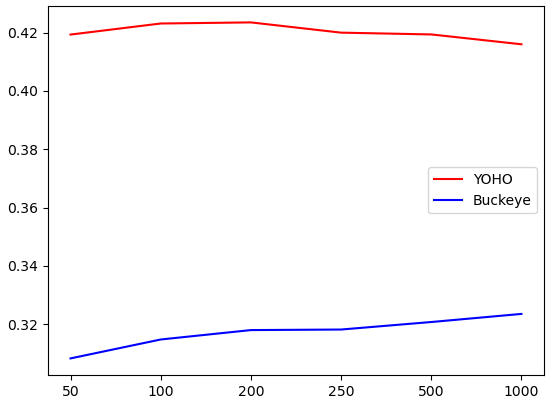}
  \caption{A plot of the word segmentation (F1-score) as a function of the number of training samples used for YOHO and Buckeye. YOHO segmentation does not benefit from large dataset due to its small vocabulary size. Buckeye experiences some accuracy gains with increasing training set size due to its large vocabulary.}
  \label{fig:train_size}
\end{figure}

\textbf{Training set size for kNN.} We investigated the effect of the number of training set utterances on the segmentation accuracy in Fig.~\ref{fig:train_size}. We observed that even a small number of utterances provides strong performance. For Buckeye there is some improvement for larger training sets, whereas for YOHO there is no improvement. This is expected as larger vocabulary datasets will naturally require larger training datasets. For both datasets, even a training set of $200$ provides sufficiently good results.

\begin{table}[th]
  \caption{Word segmentation accuracy ablation on the YOHO validation set. We can observe that a window size of duration $200$ ms ($win$ = 10) achieves the best results. The number of kNN is immaterial once it exceeds a certain threshold ($\geq 5$). Concatenating features across the window is superior to average pooling.}
  \label{tab:segknn_results_params}
  \centering
  \resizebox{\columnwidth}{!}{
  \begin{tabular}{l c c c c c c}
    \toprule
   \multicolumn{1}{c}{} & \multicolumn{1}{c}{\textbf{k}} &  \multicolumn{1}{c}{\textbf{win}} &  \multicolumn{1}{c}{\textbf{concat/avg}} & \multicolumn{1}{c}{\textbf{Prec.}} & \multicolumn{1}{c}{\textbf{Recall}} & \multicolumn{1}{c}{\textbf{F-score}}  \\
    \midrule
     &   10   &  5   & concat   &  17.1   &  24.0   & 20.0 \\
     &   10   & 10    &  concat  &  39.4   &  43.7  & 41.4 \\
     &   10   &   15  &  concat  &  36.2   & 39.2   & 37.6\\
    \midrule
     &    1  &  10   & concat   & 32.1   & 38.2    & 34.9\\
     &    5  &   10  &  concat  &  39.5   &  44.2  & 41.7 \\
     &   10   & 10    &  concat  &  39.4   &  43.7  & 41.4 \\
     &   20   &   10  &   concat &  \bf{40.8}   &  \bf{45.0}   & \bf{42.8} \\
     &   30   &   10  &   concat &  39.8    &  44.5    & 42.0  \\
     \midrule
     &     5  &   10  &   avg &    25.0  &  30.3     &  27.4  \\
     &     10  &   10  &   avg &  27.9    &   33.2    &  30.3  \\
     &     20  &   10  &   avg &  28.5    &   33.8    &  30.9  \\
     \bottomrule
  \end{tabular}}
\end{table}

\section{Discussion}

\noindent\textbf{Anomaly detection and word segmentation.} We demonstrated that anomaly detection techniques are helpful for word segmentation. Deep anomaly detection techniques extend beyond density estimation, in fact, some of the most successful techniques \cite{golan2018deep, tack2020csi} use out-of-distribution generalization performance on auxiliary tasks for anomaly scoring. Extending these to word segmentation is a promising future direction.  

\noindent\textbf{Feature adaptation.} One limitation of this work is that density estimation was performed on pretrained features directly without adaptation. Previous work (e.g. \cite{reiss2021panda}) has shown that feature adaptation can improve anomaly detection performance. Future work should examine unsupervised feature adaptation for word segmentation. Note however that standard techniques e.g. \cite{reiss2021panda} cannot directly be used as it assumes a unimodal distribution of inlier data which is invalid in this case.

\noindent\textbf{Alternative density estimators.} Here, we made use of the kNN density estimator. This estimator is very powerful as it is non-parametric and therefore effective on arbitrary inlier distributions. It requires virtually no training and has few hyperparameters. On the other hand, the kNN estimator has significant limitations including: high memory requirements, slow inference speed and potentially high sample complexity. Note that the analysis in Sec.~\ref{subsection:ablation} showed this sample complexity not to be a serious issue here. Another popular choice for density estimation is approximating the inlier distribution as a multi-dimensional Gaussian distribution \cite{lee2018simple}. This is however of limited effectiveness when the true distribution is not Gaussian, as is the case here. Future work may consider Gaussian Mixture Model density estimators \cite{glodek2013ensemble} which enjoy the benefits of parametric density estimators while dealing with multimodality. Alternatively, recent work on deep flow-base density models \cite{yu2021fastflow} may be used - as they can potentially improve beyond Gaussian models (at the expense of significantly increasing training times).

\noindent\textbf{Phoneme discovery.} Our method can achieve good word segmentation accuracy without the requirement for a preliminary phoneme discovery stage. However, in some cases, phoneme discovery is in fact the task of interest. Although not the objective of this paper, we expect that our method can be adapted to the phone segmentation setting. The core idea i.e. that segments containing a sub-sequence of a single term are more frequent than segments containing sub-sequences from multiple terms should be extensible to phonemes. Evidence for this hypothesis is provided by the success of contrastive phoneme segmentation methods e.g. \cite{kreuk2020self, bhati2021segmental}. To adapt to this setting, we expect that a smaller window size and a lower layer of the neural network may achieve the best performance. We leave this investigation for future work. 

\section{Conclusions}

We presented a simple, single-stage approach for unsupervised word segmentation. Our method, DSegkNN is based on anomaly detection using strong self-supervised pretrained features. Despite the simplicity of the method, DSegKNN achieved outstanding performance, which is comparable to the top existing methods while not requiring phoneme discovery.

\bibliographystyle{IEEEtran}
\bibliography{mybib}

\begin{thebibliography}{10}
\providecommand{\url}[1]{#1}
\csname url@samestyle\endcsname
\providecommand{\newblock}{\relax}
\providecommand{\bibinfo}[2]{#2}
\providecommand{\BIBentrySTDinterwordspacing}{\spaceskip=0pt\relax}
\providecommand{\BIBentryALTinterwordstretchfactor}{4}
\providecommand{\BIBentryALTinterwordspacing}{\spaceskip=\fontdimen2\font plus
\BIBentryALTinterwordstretchfactor\fontdimen3\font minus
  \fontdimen4\font\relax}
\providecommand{\BIBforeignlanguage}[2]{{%
\expandafter\ifx\csname l@#1\endcsname\relax
\typeout{** WARNING: IEEEtran.bst: No hyphenation pattern has been}%
\typeout{** loaded for the language `#1'. Using the pattern for}%
\typeout{** the default language instead.}%
\else
\language=\csname l@#1\endcsname
\fi
#2}}
\providecommand{\BIBdecl}{\relax}
\BIBdecl

\bibitem{jansen2013summary}
A.~Jansen, E.~Dupoux, S.~Goldwater, M.~Johnson, S.~Khudanpur, K.~Church,
  N.~Feldman, H.~Hermansky, F.~Metze, R.~Rose \emph{et~al.}, ``A summary of the
  2012 jhu clsp workshop on zero resource speech technologies and models of
  early language acquisition,'' in \emph{2013 IEEE International Conference on
  Acoustics, Speech and Signal Processing}.\hskip 1em plus 0.5em minus
  0.4em\relax IEEE, 2013, pp. 8111--8115.

\bibitem{rasanen2012computational}
O.~R{\"a}s{\"a}nen, ``Computational modeling of phonetic and lexical learning
  in early language acquisition: Existing models and future directions,''
  \emph{Speech Communication}, vol.~54, no.~9, pp. 975--997, 2012.

\bibitem{dupoux2018cognitive}
E.~Dupoux, ``Cognitive science in the era of artificial intelligence: A roadmap
  for reverse-engineering the infant language-learner,'' \emph{Cognition}, vol.
  173, pp. 43--59, 2018.

\bibitem{kamper2017embedded}
H.~Kamper, K.~Livescu, and S.~Goldwater, ``An embedded segmental k-means model
  for unsupervised segmentation and clustering of speech,'' in \emph{2017 IEEE
  Automatic Speech Recognition and Understanding Workshop (ASRU)}.\hskip 1em
  plus 0.5em minus 0.4em\relax IEEE, 2017, pp. 719--726.

\bibitem{kamper2017segmental}
H.~Kamper, A.~Jansen, and S.~Goldwater, ``A segmental framework for
  fully-unsupervised large-vocabulary speech recognition,'' \emph{Computer
  Speech \& Language}, vol.~46, pp. 154--174, 2017.

\bibitem{shain2020acquiring}
C.~Shain and M.~Elsner, ``Acquiring language from speech by learning to
  remember and predict,'' in \emph{Proceedings of the 24th Conference on
  Computational Natural Language Learning}, 2020, pp. 195--214.

\bibitem{rasanen2020unsupervised}
O.~R{\"a}s{\"a}nen and M.~A.~C. Bland{\'o}n, ``Unsupervised discovery of
  recurring speech patterns using probabilistic adaptive metrics,'' \emph{arXiv
  preprint arXiv:2008.00731}, 2020.

\bibitem{kamper2020towards}
H.~Kamper and B.~van Niekerk, ``Towards unsupervised phone and word
  segmentation using self-supervised vector-quantized neural networks,''
  \emph{arXiv preprint arXiv:2012.07551}, 2020.

\bibitem{bhati2021segmental}
S.~Bhati, J.~Villalba, P.~{\.Z}elasko, L.~Moro-Velazquez, and N.~Dehak,
  ``Segmental contrastive predictive coding for unsupervised word
  segmentation,'' \emph{arXiv preprint arXiv:2106.02170}, 2021.

\bibitem{johnson2009improving}
M.~Johnson and S.~Goldwater, ``Improving nonparameteric bayesian inference:
  experiments on unsupervised word segmentation with adaptor grammars,'' in
  \emph{Proceedings of human language technologies: The 2009 annual conference
  of the north American chapter of the association for computational
  linguistics}, 2009, pp. 317--325.

\bibitem{cuervo2021contrastive}
S.~Cuervo, M.~Grabias, J.~Chorowski, G.~Ciesielski, A.~{\L}a{\'n}cucki,
  P.~Rychlikowski, and R.~Marxer, ``Contrastive prediction strategies for
  unsupervised segmentation and categorization of phonemes and words,''
  \emph{arXiv preprint arXiv:2110.15909}, 2021.

\bibitem{reiss2021panda}
T.~Reiss, N.~Cohen, L.~Bergman, and Y.~Hoshen, ``Panda: Adapting pretrained
  features for anomaly detection and segmentation,'' in \emph{Proceedings of
  the IEEE/CVF Conference on Computer Vision and Pattern Recognition}, 2021,
  pp. 2806--2814.

\bibitem{campbell1995testing}
J.~P. Campbell, ``Testing with the yoho cd-rom voice verification corpus,'' in
  \emph{1995 international conference on acoustics, speech, and signal
  processing}, vol.~1.\hskip 1em plus 0.5em minus 0.4em\relax IEEE, 1995, pp.
  341--344.

\bibitem{pitt2005buckeye}
M.~A. Pitt, K.~Johnson, E.~Hume, S.~Kiesling, and W.~Raymond, ``The buckeye
  corpus of conversational speech: Labeling conventions and a test of
  transcriber reliability,'' \emph{Speech Communication}, vol.~45, no.~1, pp.
  89--95, 2005.

\bibitem{kuhn1955hungarian}
H.~W. Kuhn, ``The hungarian method for the assignment problem,'' \emph{Naval
  research logistics quarterly}, vol.~2, no. 1-2, pp. 83--97, 1955.

\bibitem{munkres1957algorithms}
J.~Munkres, ``Algorithms for the assignment and transportation problems,''
  \emph{Journal of the society for industrial and applied mathematics}, vol.~5,
  no.~1, pp. 32--38, 1957.

\bibitem{kreuk2020self}
F.~Kreuk, J.~Keshet, and Y.~Adi, ``Self-supervised contrastive learning for
  unsupervised phoneme segmentation,'' \emph{arXiv preprint arXiv:2007.13465},
  2020.

\bibitem{mcauliffe2017montreal}
M.~McAuliffe, M.~Socolof, S.~Mihuc, M.~Wagner, and M.~Sonderegger, ``Montreal
  forced aligner: Trainable text-speech alignment using kaldi.'' in
  \emph{Interspeech}, 2017, pp. 498--502.

\bibitem{rasanen2009improved}
O.~J. R{\"a}s{\"a}nen, U.~K. Laine, and T.~Altosaar, ``An improved speech
  segmentation quality measure: the r-value,'' in \emph{Tenth Annual Conference
  of the International Speech Communication Association}.\hskip 1em plus 0.5em
  minus 0.4em\relax Citeseer, 2009.

\bibitem{baevski2020wav2vec}
A.~Baevski, Y.~Zhou, A.~Mohamed, and M.~Auli, ``wav2vec 2.0: A framework for
  self-supervised learning of speech representations,'' \emph{Advances in
  Neural Information Processing Systems}, vol.~33, pp. 12\,449--12\,460, 2020.

\bibitem{hsu2021hubert}
W.-N. Hsu, B.~Bolte, Y.-H.~H. Tsai, K.~Lakhotia, R.~Salakhutdinov, and
  A.~Mohamed, ``Hubert: Self-supervised speech representation learning by
  masked prediction of hidden units,'' \emph{IEEE/ACM Transactions on Audio,
  Speech, and Language Processing}, vol.~29, pp. 3451--3460, 2021.

\bibitem{golan2018deep}
I.~Golan and R.~El-Yaniv, ``Deep anomaly detection using geometric
  transformations,'' in \emph{NeurIPS}, 2018.

\bibitem{tack2020csi}
J.~Tack, S.~Mo, J.~Jeong, and J.~Shin, ``Csi: Novelty detection via contrastive
  learning on distributionally shifted instances,'' \emph{NeurIPS}, 2020.

\bibitem{lee2018simple}
K.~Lee, K.~Lee, H.~Lee, and J.~Shin, ``A simple unified framework for detecting
  out-of-distribution samples and adversarial attacks,'' \emph{Advances in
  neural information processing systems}, vol.~31, 2018.

\bibitem{glodek2013ensemble}
M.~Glodek, M.~Schels, and F.~Schwenker, ``Ensemble gaussian mixture models for
  probability density estimation,'' \emph{Computational Statistics}, vol.~28,
  no.~1, pp. 127--138, 2013.

\bibitem{yu2021fastflow}
J.~Yu, Y.~Zheng, X.~Wang, W.~Li, Y.~Wu, R.~Zhao, and L.~Wu, ``Fastflow:
  Unsupervised anomaly detection and localization via 2d normalizing flows,''
  \emph{arXiv preprint arXiv:2111.07677}, 2021.

\end{thebibliography}

\end{document}